\documentclass[final,5p,times,twocolumn]{elsarticle}

\usepackage{amssymb}
\usepackage{amsmath}
\usepackage{bm}

\journal{Journal of Magnetism and Magnetic Materials}

\begin{document}

\begin{frontmatter}



\title{Fermi-Surface Topology and Pairing Symmetry in BiS$_2$-Based Layered Superconductors}

\author{Tomoaki Agatsuma}
\author{Takashi Hotta\corref{cor}}\ead{hotta@tmu.ac.jp}
\address{Department of Physics, Tokyo Metropolitan University,
1-1 Minami-Osawa, Hachioji, Tokyo 192-0397, Japan}
\cortext[cor]{Corresponding author}

\begin{abstract}
In order to clarify superconducting properties of
BiS$_2$-based layered superconductors LaO$_{1-x}$F$_x$BiS$_2$,
we evaluate the superconducting transition temperature $T_{\rm c}$
and analyze the gap function in a weak-coupling limit
on the basis of the two-dimensional two-orbital Hubbard model.
It is found that $T_{\rm c}$ becomes maximum at $x=0.6$ in the present calculations.
For $x<0.45$, we obtain the $d$-wave gap of which line nodes do not cross
the pocket-like Fermi-surface curves with the centers at X and Y points.
On the other hand, for $x>0.45$, the extended $s$-wave gap is found
for a couple of large Fermi-surface curves with the centers at M and X points.
The variation of the Fermi-surface topology plays a key role
for the gap symmetry in BiS$_2$-based superconductors.
\end{abstract}


\end{frontmatter}


\section{Introduction}

Since the discovery of new BiS$_2$-based layered superconductors
LaO$_{1-x}$F$_x$BiS$_2$
by Mizuguchi {\it et al}.\cite{Mizuguchi1,Mizuguchi2,Mizuguchi3},
both experimental and theoretical investigations for the normal and
superconducting properties on BiS$_2$-based materials
have been performed intensively.
This material group has the characteristic layered structure,
similar to high-$T_{\rm c}$ cuprate superconductors,
where $T_{\rm c}$ denotes the superconducting transition temperature.
Namely, BiS$_2$ superconducting layer is sandwiched
by insulating block layers.

Among several BiS$_2$-based compounds,
first we pick up the mother compound LaOBiS$_2$.
This material is insulating, but by substituting F for O,
we can dope electrons into the BiS$_2$ layer.
Then, the system becomes metallic and at low temperatures,
superconductivity occurs eventually.
In particular, at $x=0.5$, the superconducting transition temperature
$T_{\rm c}$ becomes maximum \cite{Mizuguchi3}.
The highest $T_{\rm c}$ among BiS$_2$ material groups
has been obtained in the sample synthesized
under high pressures \cite{Mizuguchi4}.
It has been reported that the onset $T_{\rm c}$ is 11.1 K
and the temperature at which the resistivity becomes zero is 8.5 K.

Although the mechanism of superconductivity in BiS$_2$-based materials
has not been confirmed yet,
the canonical model for this system has been proposed
by Usui {\it et al}.\cite{Usui,Kuroki},
just after the discovery of BiS$_2$-based superconductors.
The minimal model to describe electronic structure of BiS$_2$ layer is
the two-band Hamiltonian composed of
Bi 6$p_x$ and 6$p_y$ orbitals on the two-dimensional square lattice.
The appearance of superconductivity has been discussed from various
theoretical viewpoints on the basis of this two-orbital Hubbard model 
\cite{Martins,Zhou,Yang,Liang,Wu}.
The effects of electron-phonon interaction \cite{Yildirim,Wan,Li}
and spin-orbit coupling \cite{Gao} have been also discussed.
The relation between the characteristic change of the
Fermi-surface topology and the symmetry of the superconducting gap function
have been pointed out in previous works \cite{Usui,Kuroki,Martins,Zhou,Yang,Liang}.

Concerning the gap symmetry,
some theoretical studies suggested that the gap function in
BiS$_2$-based layered superconductors could be
explained by extended $s$-wave, $d$-wave, triplet or
other unconventional paring scenarios \cite{Usui,Martins,Zhou,Yang,Liang,Wu}.
On the other hand, $s$-wave pairing due to electron-phonon interaction
has been also discussed \cite{Yildirim,Wan,Li}.
From experimental viewpoints,
the results consistent with $s$- or extended $s$-wave pairing
have been found in the temperature dependence of the superfluid density
in Bi$_4$O$_4$S$_3$ \cite{Srivastava},
LaO$_{0.5}$F$_{0.5}$BiS$_2$ \cite{Lamura},
and NdO$_{1-x}$F$_{x}$BiS$_2$ \cite{Jiao}.
In particular, for NdO$_{1-x}$F$_{x}$BiS$_2$, it has been reported that
the gap symmetry is consistent with
$s$- or extended $s$-wave both for $x=0.3$ and $0.5$ \cite{Jiao}.
It is interesting to consider a way to reconcile those results on the symmetry of the gap
function from a viewpoint
of Fermi-surface topology characteristic to multi-band systems.

In this paper, we analyze the two-dimensional two-band model
including intra- and inter-orbital Coulomb interactions
within a random phase approximation (RPA) to derive the effective pairing
interaction to induce the Cooper pair.
Then, we solve the gap equation to obtain $T_{\rm c}$ and analyze
the gap symmetry by focusing on the Fermi-surface topology.
For $x=0.3$ and $0.4$, we obtain the $d$-wave gap, but the disconnected
Fermi-surface curves with the centers at X and Y points do $not$ cross
the line nodes of the $d$-wave gap.
When we increase $x$, we observe the elevation of $T_{\rm c}$
except for a narrow region around at $x=0.45$ and
the conversion of the gap symmetry from $d$-wave to extended $s$-wave.
In fact, $T_{\rm c}$ takes the maximum value at $x=0.6$.
For $x=0.5$ and $0.6$, we obtain the extended $s$-wave gap 
for a couple of large Fermi-surface curves
with the centers at $\Gamma$ and M points.
It is proposed that the conversion of the gap symmetry is
characteristically induced by the variation of the Fermi-surface
topology in the BiS$_2$-layered materials.
Throughout this paper, we use such units as $\hbar=k_{\rm B}=1$.

\section{Model and Formulation}

\subsection{Model Hamiltonian}

The model Hamiltonian is given by~\cite{Usui,Kuroki}
\begin{eqnarray}
  H &=& \sum_{\bm{k}\sigma\tau\tau'}
    \varepsilon_{\bm{k}\tau\tau'}
    c_{\bm{k}\tau\sigma}^{\dag}c_{\bm{k}\tau'\sigma}
    +U \sum_{\bm{i} \tau}
    n_{\bm{i} \tau \uparrow} n_{\bm{i} \tau \downarrow}
    \nonumber\\
    &+&U^{\prime} \sum_{\bm{i}}
    n_{\bm{i}1} n_{\bm{i}2}
    + J \sum_{\bm{i},\sigma,\sigma^{\prime}}
    c^{\dagger}_{\bm{i} 1 \sigma}
    c^{\dagger}_{\bm{i} 2 \sigma^{\prime}}
    c_{\bm{i} 1 \sigma^{\prime}}
    c_{\bm{i} 2 \sigma} \\
    &+& J^{\prime}\sum_{\bm{1},\tau \ne \tau^{\prime}}
    c^{\dagger}_{\bm{i} \tau \uparrow}
    c^{\dagger}_{\bm{i} \tau \downarrow}
    c_{\bm{i} \tau^{\prime} \downarrow}
    c_{\bm{i} \tau^{\prime} \uparrow},\nonumber
\end{eqnarray}
where $c_{\bm{k}\mu\sigma}$ is the annihilation operator of electron with
wave-vector $\bm{k}$ and spin $\sigma$ in the orbital $\tau$ (=1 and 2),
$\varepsilon_{\bm{k}\tau\tau'}$ is the electron energy
between $\tau$ and $\tau'$ orbitals,
$n_{\bm{i} \tau \sigma}=c^{\dagger}_{\bm{i} \tau \sigma} c_{\bm{i} \tau \sigma}$,
$n_{\bm{i} \tau}=\sum_{\sigma} n_{\bm{i} \tau \sigma}$,
and the coupling constants $U$, $U^{\prime}$, $J$, and $J^{\prime}$
denote the intra-orbital Coulomb, inter-orbital Coulomb, exchange,
and pair-hopping interactions, respectively.
Note the relations of $U=U'+J+J'$ and $J=J'$.

\subsection{Electron dispersion}

As for the electron dispersion relation, by following the results
in the paper of Usui {\it et al}.\cite{Usui}, we obtain
\begin{eqnarray}
\varepsilon_{\bm{k}11} &=&
t_0+2t_1 (\cos k_x+\cos k_y) \nonumber \\
&+& 2t_3 \cos (k_x-k_y) +2t_4 \cos (k_x+k_y)  \nonumber \\
&+& 2t_6 [\cos (2k_x+k_y) + \cos (k_x+2k_y)] \nonumber\\
&+& 2t_8  [\cos (2k_x-k_y) + \cos (k_x-2 k_y)],
\end{eqnarray}
\begin{eqnarray}
\varepsilon_{\bm{k}22} &=&
t_0+2t_1 (\cos k_x+\cos k_y) \nonumber \\
&+& 2t_3 \cos (k_x+k_y) +2t_4 \cos (k_x-k_y)  \nonumber \\
&+& 2t_6 [\cos (2k_x-k_y) + \cos (k_x-2k_y)] \nonumber\\
&+& 2t_8  [\cos (2k_x+k_y) + \cos (k_x+2 k_y)],
\end{eqnarray}
and
\begin{eqnarray}
\varepsilon_{\bm{k}12} &=&
\varepsilon_{\bm{k}21} \nonumber \\
&=&2t_2 (\cos k_x-\cos k_y)+2t_5(\cos 2k_x -\cos 2k_y) \nonumber \\
&+&4t_7 (\cos 2k_x \cos k_y - \cos k_x \cos 2k_y).
\end{eqnarray}
The values of $t_0$$\sim$$t_8$ are given by
$t_0$=$2.811$, $t_1$=$-0.167$, $t_2$=$0.107$,
$t_3$=$0.880$, $t_4$=$0.094$, $t_5$=$-0.028$,
$t_6$=$0.014$, $t_7$=$0.020$, and $t_8$=$0.069$
in the unit of eV \cite{Usui}.

The band dispersion is given by
\begin{equation}
  E_{{\bm k}\pm} = \frac{1}{2}
   \left[ \varepsilon_{{\bm k}11}+
\varepsilon_{{\bm k}22}
\pm \sqrt{ (\varepsilon_{{\bm k}11}-
\varepsilon_{{\bm k}22})^2+
\varepsilon_{{\bm k}12}^2} \right].
\end{equation}
Hereafter we define $E_{{\bm k}+}=E_{{\bm k}A}$
and $E_{{\bm k}-}=E_{{\bm k}B}$,
which are called the A and B bands, respectively.

The band structure has been discussed in Ref.~\cite{Usui},
but in order to make this paper self-contained,
we show $E_{{\bm k}A}$ and $E_{{\bm k}B}$ along the path
with high symmetry on the Brillouin zone in Fig.~1(a).
For $x=0.3$ and $0.4$, as shown in the insets,
we observe that the Fermi levels cross only the B band
near the X point, while between M and $\Gamma$ points, they do not
cross the band B.
For $x=0.5$, the Fermi level turns to cross the band B between M and $\Gamma$ points,
but it is still under the band A at the X point.
For $x=0.6$, the Fermi level crosses not only the B band but also the A
band near the X point.

\begin{figure}[t]
\centering
\includegraphics[width=0.98\linewidth]{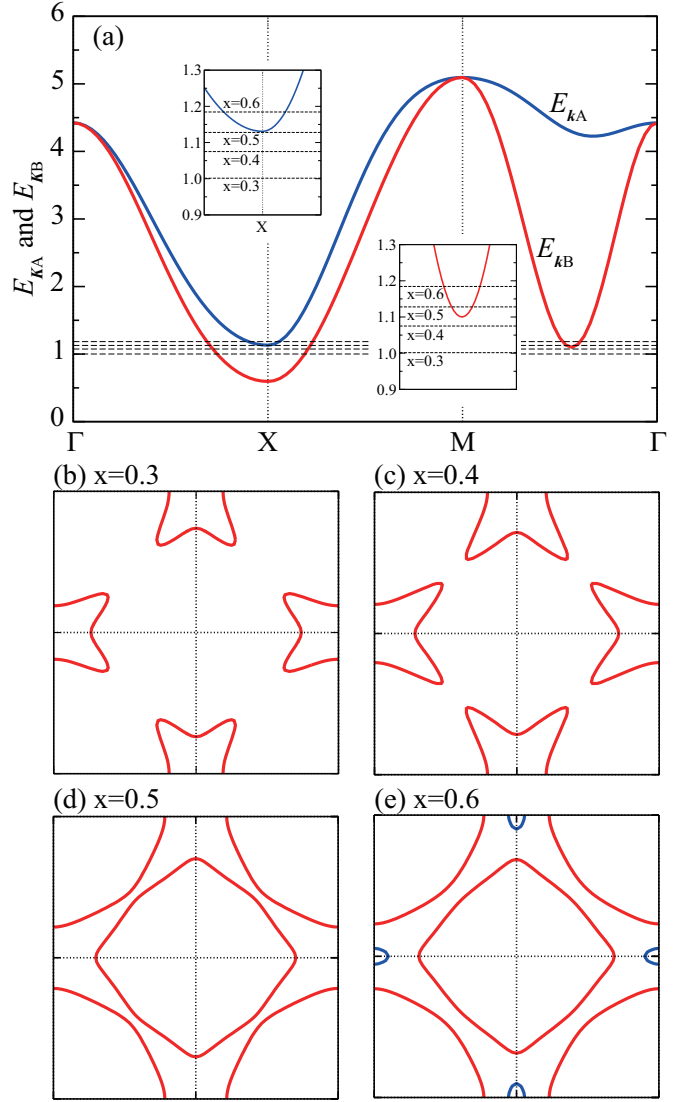}
\caption{
(a) $E_{{\bm k}A}$ and $E_{{\bm k}B}$ along the path of
$\Gamma$$\rightarrow$X$\rightarrow$M$\rightarrow$$\Gamma$.
The horizontal broken lines denote the positions of Fermi levels
for $x=0.3$, $0.4$, $0.5$ and $0.6$
from the bottom to the top.
Insets denote the band structure in magnified scales
around at $(\pi,0)$ and a point between $(\pi,\pi)$ and $(0,0)$.
Fermi-surface curves for (b) $x=0.3$, (c) $x=0.4$, (d) $x=0.5$,
and (e) $x=0.6$.
}
\end{figure}

The above characters appear in the change of the Fermi-surface structure,
as shown in Figs.~1(b)-1(e).
In Fig.~1(b), we depict the Fermi-surface curves for $x=0.3$.
A characteristic issue is that pocket-like disconnected Fermi-surface curves
appear around at the X and Y points.
We remark that these Fermi-surface curves originate from the B band and
they do not cross the lines of $k_x=\pm k_y$.
This issue will be important in the discussion of the gap function later.
Another characteristic point is the shape of the Fermi-surface curves.
We observe the disconnected Fermi-surface curves, which possess
the projections extending to the direction to those of
other Fermi-surface curves.
For $x=0.4$, we still observe pocket-like
disconnected Fermi-surface curves around at the X and Y points,
as shown in Fig.~1(c).
Note that the projections further extend to
the directions to other Fermi-surface curves, but they do not touch.

When we further increase $x$, the projections eventually
touch other Fermi-surface curves and the disconnected structure
is dissolved, as observed in Fig.~1(d) for $x=0.5$.
Then, we observe a couple of large Fermi-surface curves
with the centers at the $\Gamma$ and M points.
These two Fermi-surface curves originating from the B band
are formed by the merge of the projections of
pocket-like Fermi-surface curves in Fig.~1(c).
We should pay attention to the existence of the Fermi-surface curves
on the lines of $k_x=\pm k_y$.
We also remark that the topological change of the Fermi-surface
structure occurs around at $x=0.45$,
as already mentioned in Refs.~\cite{Martins} and \cite{Liang}.

Now we further increase the values of $x$.
The Fermi-surface curves for $x=0.6$ are shown in Fig.~1(e).
Here we observe the large Fermi-surface curves around
at the $\Gamma$ and M points, although the shape is slightly deformed
from those for $x=0.5$ in Fig.~1(d).
Note also that new pocket-like Fermi-surface curves appear around
at the X and Y points.
These curves originate from the A band,
in sharp contrast to other Fermi-surface curves.
We find that there occurs the topological change for the appearance of
the Fermi-surface curve
originating from the A band at $x=0.52$,
as already pointed out in Ref.~\cite{Liang}.

\subsection{Spin and orbital susceptibilities}

In order to investigate superconductivity around the spin and/or orbital ordered phases
\cite{Takimoto1,Takimoto2,Takimoto3,Mochizuki1,Mochizuki2,Mochizuki3,Yada,Kubo1,Kubo2},
we calculate spin and orbital susceptibilities,
$\hat{\chi}^{\rm s}({\bm q})$ and $\hat{\chi}^{\rm o}({\bm q})$, 
respectively.
Here we follow the formulation and the notations in Ref.~\cite{Takimoto2}.
Within the RPA, spin and orbital susceptibilities are given
in the 4$\times$4 matrix form as
\begin{equation}
  \label{RPAchis}
  \hat{\chi}^{\rm s}({\bm q}) =
  [\hat{1}-\hat{U}^{\rm s}\hat{\chi}({\bm q})]^{-1}
  \hat{\chi}({\bm q}),
\end{equation}
and
\begin{equation}
  \label{RPAchio}
  \hat{\chi}^{\rm o}({\bm q}) =
  [\hat{1}+\hat{U}^{\rm o}\hat{\chi}({\bm q})]^{-1}
  \hat{\chi}({\bm q}),
\end{equation}
respectively.
Here labels of row and column in the matrix appear in the order
of 11, 22, 12, and 21, which are pairs of orbital indices 1 and 2.
Note that $\hat{1}$ is the 4$\times$4 unit matrix.
The interaction matrices $\hat{U}^{\rm s}$ and $\hat{U}^{\rm o}$ are
given by
\begin{equation}
\hat{U}^{\rm s}=\left(
\begin{array}{cccc}
U & J & 0 & 0 \\
J & U & 0 & 0 \\
0 & 0 & U' & J' \\
0 & 0 & J' & U'
\end{array}
\right),
\end{equation}
and
\begin{equation}
\hat{U}^{\rm o}=\left(
\begin{array}{cccc}
U & 2U'-J & 0 & 0 \\
2U'-J & U & 0 & 0 \\
0 & 0 & 2J-U' & J' \\
0 & 0 & J' & 2J-U'
\end{array}
\right),
\end{equation}
respectively.
Each matrix element of $\hat{\chi}({\bm q})$ is defined by
\begin{equation}
\chi_{\mu\nu,\alpha\beta}({\bm q})=
-T\sum_{{\bm k},n}
G^{(0)}_{\alpha\mu}({\bm k}+{\bm q},i\omega_{n})
G^{(0)}_{\nu\beta}({\bm k},i\omega_{n}),
\end{equation}
where $T$ is a temperature,
$G^{(0)}_{\mu\nu}({\bm k},i\omega_{n})$
is the non-interacting Green's function propagating
between $\mu$- and $\nu$-orbitals,
and $\omega_n=\pi T(2n+1)$ with an integer $n$.
The instabilities for the spin- and orbital-ordered phases
are determined by the conditions of
${\rm det}[\hat{1}-\hat{U}^{\rm s}\hat{\chi}({\bm q})]$=0
and ${\rm det}[\hat{1}+\hat{U}^{\rm o}\hat{\chi}({\bm q})]$=0, 
respectively.

\subsection{Gap equation}

By using $\hat{\chi}^{\rm s}({\bm q})$ and 
$\hat{\chi}^{\rm o}({\bm q})$,
we obtain the superconducting gap equation
in a weak-coupling limit as
\begin{equation}
  \label{gap}
  {\bm \Delta}^{\xi}({\bm k})=\sum_{{\bm k'}}
    \hat{V}^{\xi}({\bm k}-{\bm k'})
    \hat{\phi}({\bm k'}){\bm \Delta}^{\xi}({\bm k'}),
\end{equation}
where ${\bm \Delta}^{\xi}({\bm k})$
=[$\Delta^{\xi}_{11}({\bm k})$, $\Delta^{\xi}_{22}({\bm k})$,
$\Delta^{\xi}_{12}({\bm k})$, $\Delta^{\xi}_{21}({\bm k}$)]$^t$
is the gap function in the vector representation
for singlet ($\xi={\rm S}$) or triplet 
($\xi={\rm T}$) pairing state.
The matrix elements of the singlet- and triplet-pairing 
potentials are, respectively, given by
\begin{equation}
  V^{\rm S}_{\alpha\beta,\mu\nu}({\bm q})=
   -\frac{3}{2}
    W^{\rm s}_{\alpha\mu,\nu\beta}({\bm q})+
    \frac{1}{2}W^{\rm o}_{\alpha\mu,\nu\beta}({\bm q})
  +U^{\rm s}_{\alpha\mu,\nu\beta},
\end{equation}
\begin{equation}
  V^{\rm T}_{\alpha\beta,\mu\nu}({\bm q})=
   \frac{1}{2}W^{\rm s}_{\alpha\mu,\nu\beta}({\bm q})+
   \frac{1}{2}W^{\rm o}_{\alpha\mu,\nu\beta}({\bm q})
   -U^{\rm s}_{\alpha\mu,\nu\beta},
\end{equation}
which are composed of spin and orbital susceptibilities as
\begin{equation}
  \hat{W}^{\rm s}({\bm q})=\hat{U}^{\rm s}
  +\hat{U}^{\rm s}\hat{\chi}^{\rm s}({\bm q})\hat{U}^{\rm s},
\end{equation}
\begin{equation}
  \hat{W}^{\rm o}({\bm q})=-\hat{U}^{\rm o}
  +\hat{U}^{\rm o}\hat{\chi}^{\rm o}({\bm q})\hat{U}^{\rm o}.
\end{equation}
The element of the pair correlation function
$\hat{\phi}({\bm k})$ is given by
\begin{equation}
\phi_{\alpha\beta,\mu\nu}({\bm k})=
T\sum_{n}
G^{(0)}_{\alpha\mu}(\bm{k},-i\omega_{n})
G^{(0)}_{\nu\beta}(-\bm{k},i\omega_{n}).
\end{equation}

We obtain $T_{\rm c}$ by solving the gap equation eq.~(\ref{gap}).
Note that $T_{\rm c}$ is defined as a temperature at which
the positive maximum eigenvalue of the gap equation becomes unity.
For the evaluation of the eigenvalue of eq.~(\ref{gap}),
we use the power method.
The first Brillouin zone is divided into $64 \times 64$ meshes and
we exploit the fast-Fourier-transformation algorithm,
in order to accelerate the sum of large amount of momenta.

As for the values of Coulomb interactions,
we note the relations of $U=U'+J+J'$ and $J=J'$,
as mentioned above.
Here we assume the value of $J$ as $J=U/6$.
Then, from the above relations, we obtain $U'=2U/3$.
In this paper, we fix the value of $U$ as $U=2$ eV.
Note that the energy unit is set as eV in this paper.
The bandwidth $W$ of the present tight-binding band structure
is $W=4.5$ eV.
Thus, the value of $U=2$ is not so large in comparison with $W$.

\section{Calculation Results}

\subsection{Phase diagram}

Let us show our calculation results for the transition temperature $T_{\rm c}$
as a function of doping $x$ in Fig.~2.
First of all, we should note the magnitude of the vertical temperature axis.
Since the energy unit is eV in the present calculations,
the maximum value of $T_{\rm c}$ becomes about 600 K,
if we simply use the present energy unit.
Here we should pay attention to the fact that 
we have not included at all the effects of the retardation of effective interactions
and the quasi-particle life-time.
If we correctly include the vertex corrections in addition to above strong-coupling effects,
$T_{\rm c}$ is considered to be strongly suppressed.
Thus, it is not necessary to consider seriously the magnitude of $T_{\rm c}$
in the present calculations.
Rather we should pay our attention to the relative change of
$T_{\rm c}$ by the doping $x$.

\begin{figure}[t]
\centering
\includegraphics[width=\linewidth]{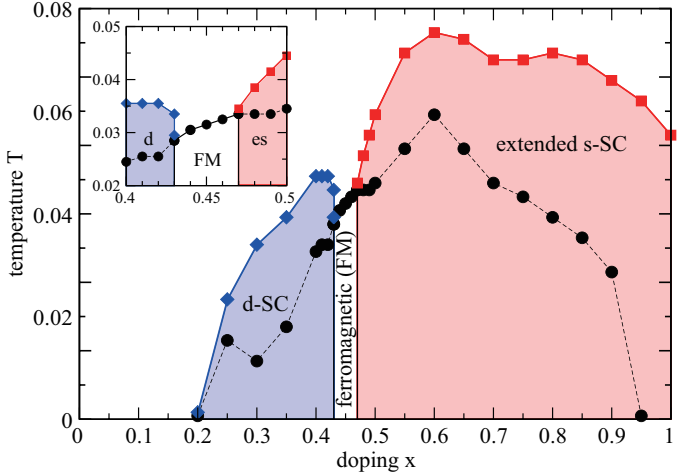}
\caption{Phase diagram on the plane of doping $x$ and temperature $T$.
Solid black circles denote the magnetic instability temperature $T_{\rm M}$,
while solid red squares and blue diamonds indicate $T_{\rm c}$
for the extended $s$- and $d$-wave symmetries, respectively.
Note that the solid curve denotes the phase boundary, but the broken curve does not.
Inset shows the part of the phase diagram around at $x=0.45$ in a magnified scale.}
\end{figure}

The lower curve denotes the magnetic instability temperature $T_{\rm M}$
determined from ${\rm det}[\hat{1}-\hat{U}^{\rm s}\hat{\chi}({\bm q})]=0$.
In the present calculations, when we decrease the temperature,
the spin susceptibility always diverges prior to the orbital one
and we do not consider the orbital susceptibility.
Then, we solve the gap equation in the range of $T>T_{\rm M}$ to
consider the superconductivity induced by magnetic fluctuations
in the vicinity of the magnetic instability temperature.
Note that we cannot discuss the coexistence of magnetism and superconductivity
in the present RPA.

In Fig.~2, we immediately notice that the superconducting phase
in the weak-coupling limit appears in the wide range of the values
of doping $x$ and temperature $T$.
Roughly speaking, the shape of the curve for $T_{\rm c}$ is
similar to that for $T_{\rm M}$.
In particular, the position of the maximum $T_{\rm c}$ is exactly the same
as that of $T_{\rm M}$.
We also point out that the superconducting phase is divided into two regions:
The $d$-wave phase for $x<0.45$ and the extended $s$-wave phase for $x>0.45$.
The figure around at $x \approx 0.45$ in a magnified scale is shown in the inset of Fig.~2.
In the present calculations, in a very narrow region around at $x=0.45$,
$T_{\rm c}$ is suppressed.
Instead, we find the ferromagnetic (FM) phase in this region,
since the ordering vector is found to be ${\bm q}=(0,0)$ in this region.
Near $x=1.0$, $T_{\rm M}$ becomes zero since both $U$ and
the density of states at the Fermi level are small in this case,
but $T_{\rm c}$ keeps a large value,
although it is smaller than the peak value at $x=0.6$.
For $x<0.2$, we cannot obtain both magnetic and superconducting phases
in the RPA for the system with small electron densities.

We note that there is not significant structure in $T_{\rm c}$
around at $x=0.52$, at which the Fermi-surface topology is
changed due to the appearance of the Fermi-surface curves
originating from the A band.
We deduce that the effect of the pairing in the A band will not
be so significant,
since the superconductivity occurs mainly in the B band.
In the $d$-wave phase, except for the narrow region around at $x=0.45$,
$T_{\rm c}$ is increased with the increase of $x$.
On the other hand, in the extended $s$-wave phase, $T_{\rm c}$ takes
a broad peak at $x=0.6$.
Note that the magnetic instability curve also exhibits the peak at $x=0.6$,
but the peak structure is rather sharp.

Let us now compare our phase diagram with experimental results.
First we briefly review the experimental facts.
For LaO$_{1-x}$F$_x$BiS$_2$,
the doping dependence of $T_{\rm c}$ seems to be sensitive to
the way of sample synthesis \cite{Mizuguchi4,Deguchi,Higashinaka}.
For as-grown samples, $T_{\rm c}$ is increased between $x=0.2$ and $0.5$,
whereas it becomes almost constant between $x=0.5$ and $0.7$.
For annealed samples under high pressures,
$T_{\rm c}$ is totally increased and it becomes maximum at $x=0.5$.
In addition, $T_{\rm c}$ seems to be abruptly increased
between $x=0.4$ and $0.5$.

For NdO$_{1-x}$F$_x$BiS$_2$, we have found weak
doping dependence of $T_{\rm c}$ \cite{Demura}.
It has been found that superconductivity appears in the region of
$0.1 \le x \le 0.7$ and $T_{\rm c}$ becomes maximum at $x=0.4$,
although the change of $T_{\rm c}$ by doping $x$ is not so significant
in comparison with LaO$_{1-x}$F$_x$BiS$_2$.

In our phase diagram, the maximum $T_{\rm c}$ is found at $x=0.6$,
which is deviated from $x=0.5$ for LaO$_{1-x}$F$_x$BiS$_2$
and $x=0.4$ for NdO$_{1-x}$F$_x$BiS$_2$.
This deviation may be explained by the strong coupling calculation
or by the further inclusion of electron-phonon interaction.
These points will be discussed in future.
We remark a point that the nominal value of $x$ may be deviated
from the amount of electron doping in BiS$_2$ layer \cite{Mizuguchi5}.
We should pay due attention to this point in the comparison with
experiments. It is another future task.

Around at $x=0.45$, the Fermi-surface topology is changed
and the singlet superconductivity has been found to be suppressed.
As expected from the FM phase around at $x=0.45$,
the eigenvalue of the triplet pair has been found to be larger than
that of the singlet one in this region, although it has not
reached the unity in the present calculation.
The tendency of the triplet pair may be also related to the previous
work in Ref.~\cite{Yang}.
When we include the strong-coupling effects and the vertex corrections
in the evaluation of the normal and anomalous self-energies,
the triplet superconductivity may occur around at $x=0.45$.
This is one of interesting future issues.

\subsection{Susceptibilities and Fermi-surface curves}

In order to grasp the kind of the magnetic fluctuations for the pair formation,
we show the spin susceptibilities in the RPA
for $x=0.3$, $0.4$, $0.5$, and $0.6$ in Figs.~3 and 4
along the path of $\bm{q}=(0,0)$$\rightarrow$$(\pi,0)$
$\rightarrow$$(\pi,\pi)$$\rightarrow$$(0,0)$.
Since we have found that orbital susceptibilities are totally suppressed
and they do not show any significant structures in comparison with spin ones
in the present calculations, we show only spin susceptibilities in this paper.

\begin{figure}[t]
\centering
\includegraphics[width=\linewidth]{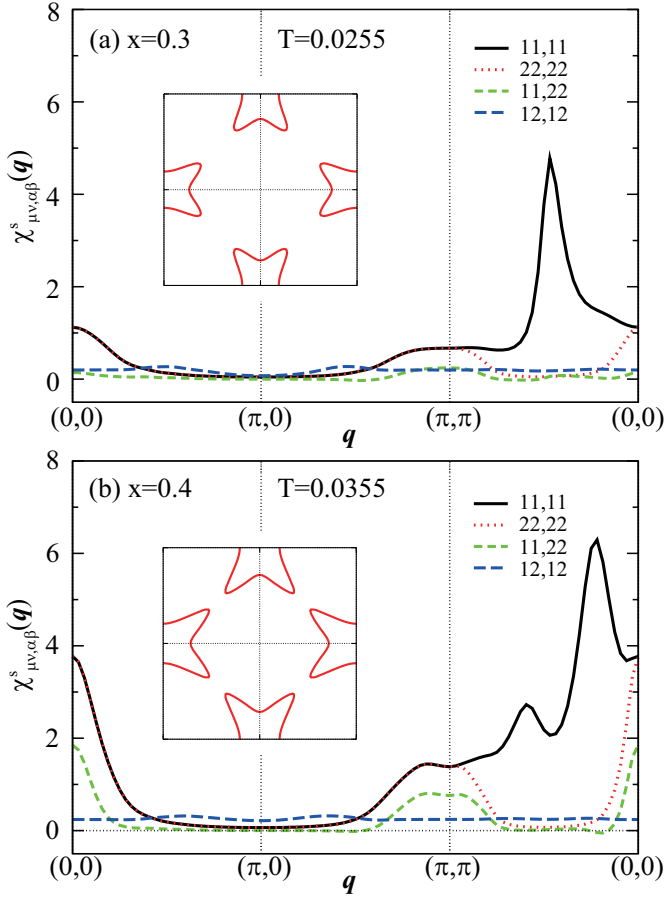}
\caption{Spin susceptibilities in the RPA
for (a) $x=0.3$ and (b) $x=0.4$.
Inset in each panel denotes the Fermi-surface curve.
}
\end{figure}

\begin{figure}[t]
\centering
\includegraphics[width=\linewidth]{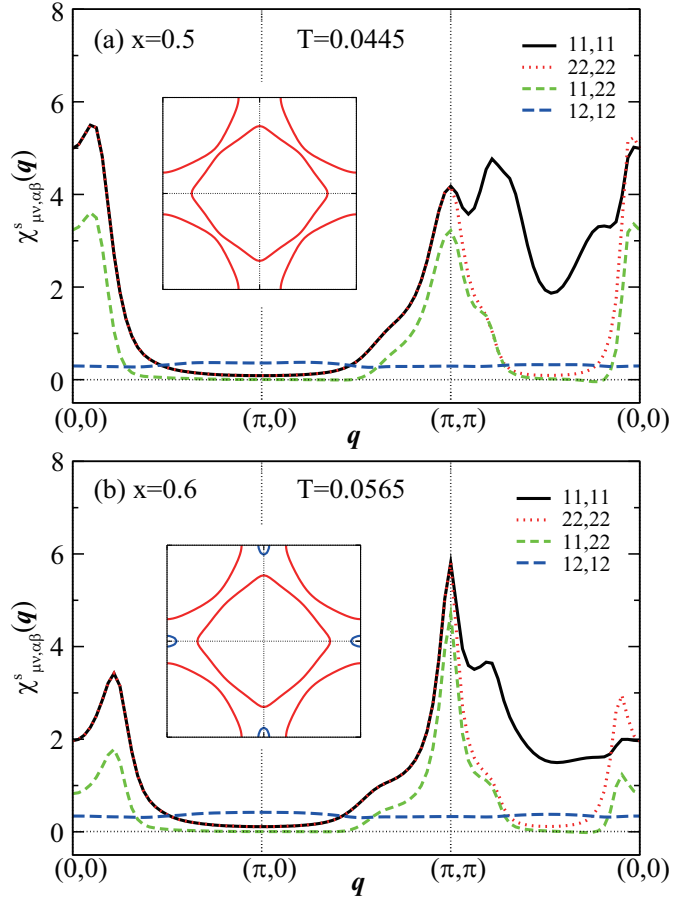}
\caption{Spin susceptibilities in the RPA
for (a) $x=0.5$ and (b) $x=0.6$.
Inset in each panel denotes the Fermi-surface curve.
}
\end{figure}

In Figs.~3(a) and 3(b), we show the spin susceptibilities,
respectively, for $x=0.3$ and $0.4$.
Note that the temperatures are set as $T_{\rm c}$,
which are $T_{\rm c}=0.0255$ for $x=0.3$ and $T_{\rm c}=0.0355$
for $x=0.4$.
We find the enhanced spin fluctuations of $\bm{q}$=$(q,q)$,
which will contribute to the Cooper-pair formation for superconductivity.
The value of $q$ becomes small when $x$ is increased from $0.3$ to $0.4$.
For $x=0.4$, we observe another small peak for large $q$.
Note that the FM fluctuations also grow for $x=0.4$
in comparison with that for $x=0.3$.
The FM fluctuations tend to be significant, when $x$ approaches $0.45$,
as expected from the existence of the FM phase around at $x=0.45$.

In Fig.~4(a) and 4(b), we exhibit the spin susceptibilities in the RPA
for $x=0.5$ and $0.6$, respectively.
Note that the temperature is set as $T=0.0445$ and $0.0565$,
which are equal to $T_{\rm c}$ at $x=0.5$ and $0.6$, respectively.
In the bare susceptibility, the maximum peak appears at $\bm{q}=(q, 0)$ with small $q$.
This peak structure is due to the nesting of a couple of large Fermi-surface curves
with the centers at the $\Gamma$ and M points \cite{Martins}.
In fact, for $x=0.5$, the peak  at $\bm{q}=(q, 0)$ is larger than other peak values
in the RPA spin susceptibility, although anti-ferromagnetic (AF) spin fluctuations also grow.
For $x=0.6$, the peak at $\bm{q}=(q, 0)$ is still significant,
but we observe a clear peak at $\bm{q}=(\pi, \pi)$,
suggesting AF spin fluctuations.
We deduce that the evolution of the AF spin fluctuations
is closely related to the elevation of $T_{\rm c}$
from $x=0.5$ to $x=0.6$.

\subsection{Gap functions and Fermi-surface curves}

Now we show the results of the gap function
for the Cooper pair from the same band,
$\Delta_{\rm A}(\bm{k})$ and $\Delta_{\rm B}(\bm{k})$,
which are obtained by the appropriate linear combinations
of $\Delta^{\rm S}_{11}({\bm k})$, $\Delta^{\rm S}_{22}({\bm k})$,
$\Delta^{\rm S}_{12}({\bm k})$, and $\Delta^{\rm S}_{21}({\bm k})$.
There exist the components of the Cooper pairs formed by different band electrons,
but we do not show them here, since their magnitudes are small
in comparison with those of the same band.
Note also  that only the singlet pair has been stabilized,
although we have investigated both singlet and triplet channels.
Thus, we show here the results for singlet gap functions.

\begin{figure}[t]
\centering
\includegraphics[width=\linewidth]{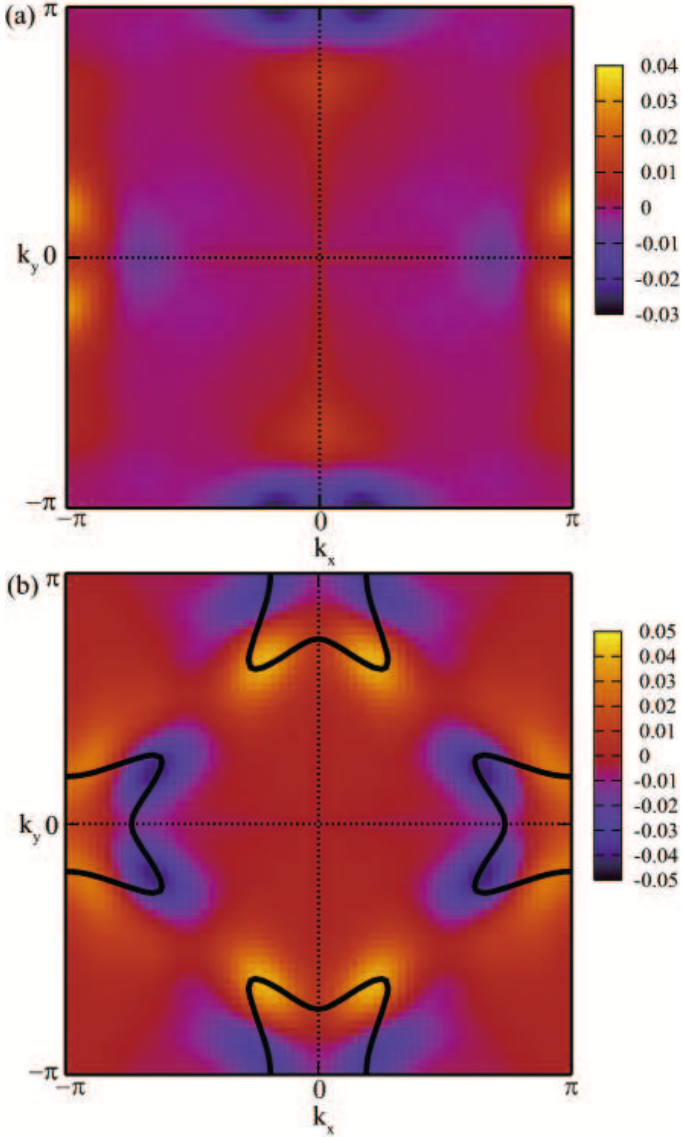}
\caption{
(a) $\Delta_{\rm A}(\bm{k})$ and (b) $\Delta_{\rm B}(\bm{k})$ for $x=0.3$.
We show the Fermi-surface curves in the figure.
}
\end{figure}

\begin{figure}[t]
\centering
\includegraphics[width=\linewidth]{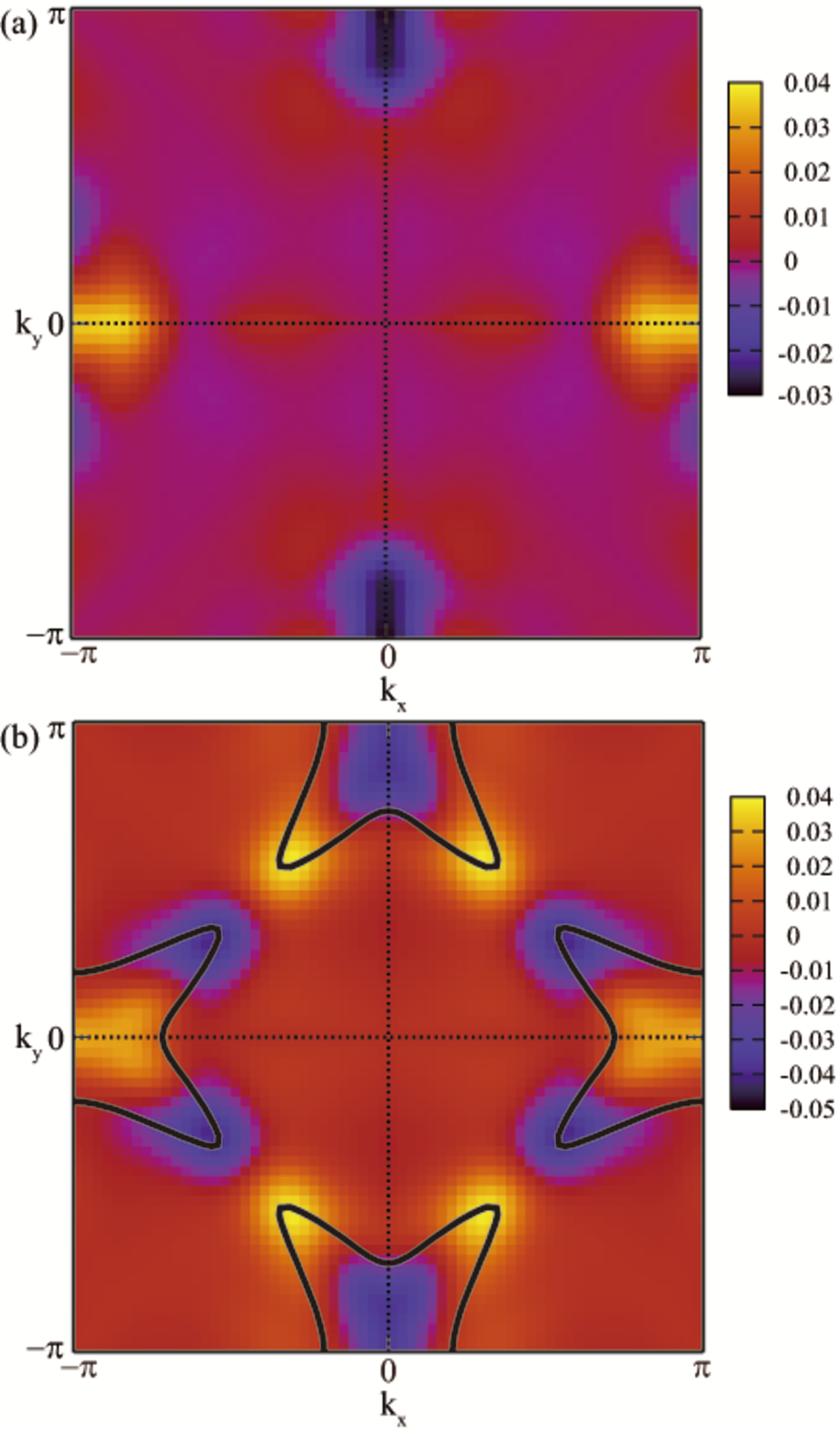}
\caption{
(a) $\Delta_{\rm A}(\bm{k})$ and (b) $\Delta_{\rm B}(\bm{k})$ for $x=0.4$.
We show the Fermi-surface curves in the figure.
}
\end{figure}

\begin{figure}[t]
\centering
\includegraphics[width=\linewidth]{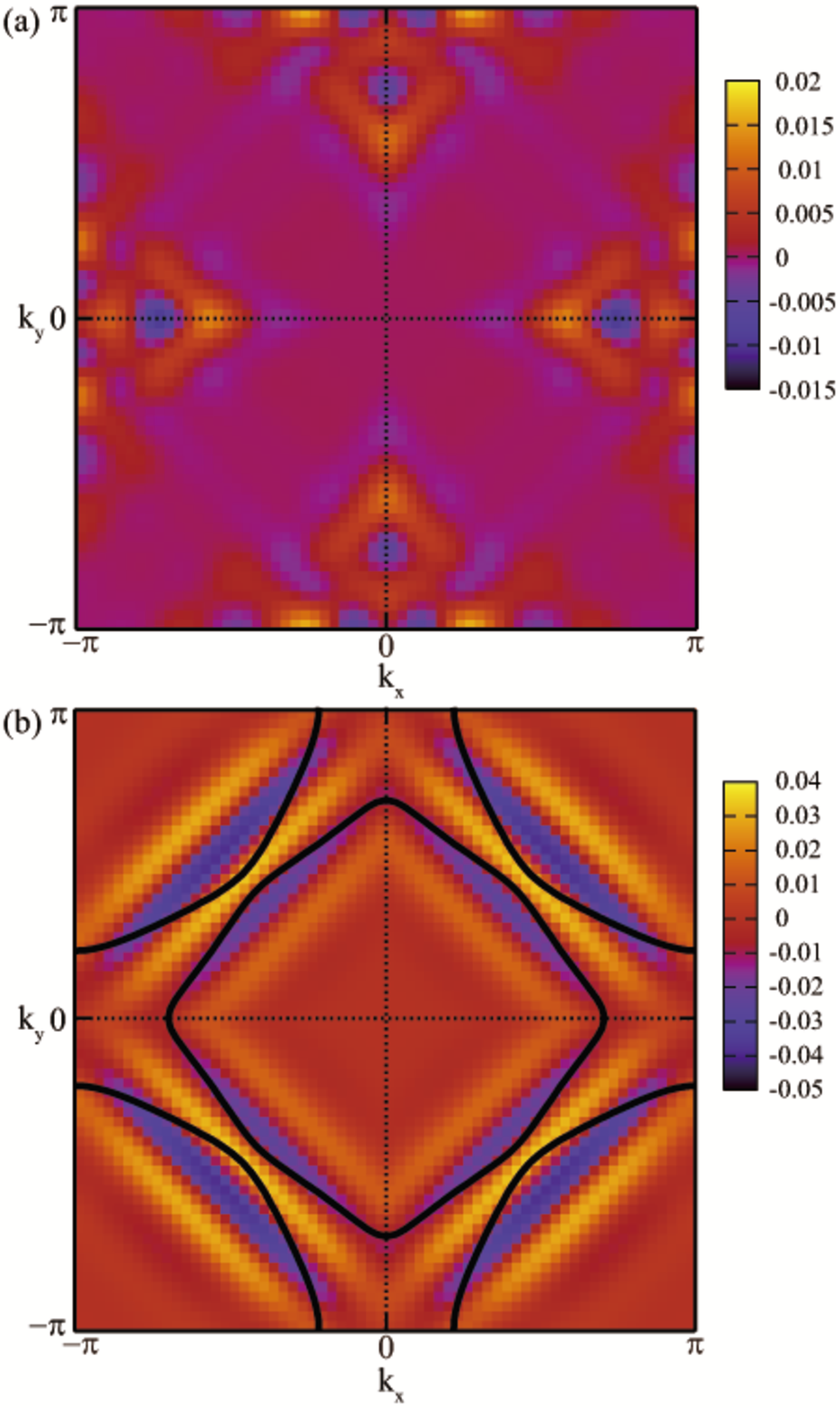}
\caption{
(a) $\Delta_{\rm A}(\bm{k})$ and (b) $\Delta_{\rm B}(\bm{k})$ for $x=0.5$.
We show the Fermi-surface curves in the figure.
}
\end{figure}

\begin{figure}[t]
\centering
\includegraphics[width=\linewidth]{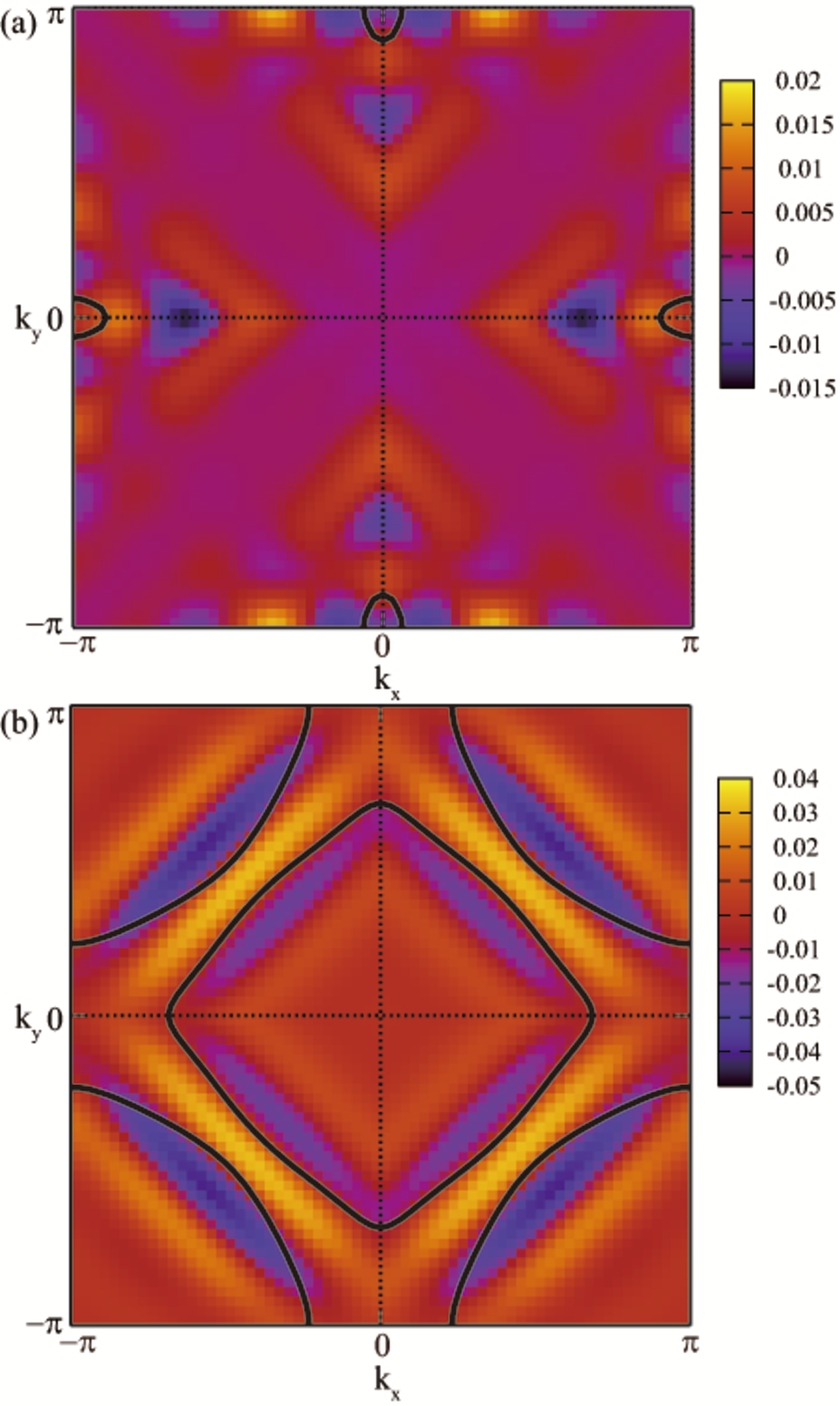}
\caption{
(a) $\Delta_{\rm A}(\bm{k})$ and (b) $\Delta_{\rm B}(\bm{k})$ for $x=0.6$.
We show the Fermi-surface curves in the figure.
}
\end{figure}

In Figs.~5, we exhibit $\Delta_{\rm A}(\bm{k})$
and $\Delta_{\rm B}(\bm{k})$ for $x=0.3$ and $T=0.0255$
by using the color gradation to visualize the momentum
dependence at a glance.
First we notice that the gap functions $\Delta_{\rm A}(\bm{k})$
and $\Delta_{\rm B}(\bm{k})$ possess the $d$-wave symmetry,
which is characterized by the sign change of the gap for the interchange
of $k_x$ and $k_y$.
In this doping, there is no Fermi-surface curve from the band A,
but we observe the significant amplitude around at the X and Y points,
since the band A is near the Fermi level on these points,
as shown in Fig.~1(a).

As for the gap function on the band B in Fig.~5(b),
we observe that the nodes of the $d$-wave, $k_x=\pm k_y$,
are running between the pocket-like Fermi-surface curves.
Namely, the nodes do not cross the Fermi-surface curves.
However, we find the sign change of the gap at the heads
of the projections for the interchange of $k_x$ and $k_y$.
Note that the signs of the gap functions near the X and Y points are
also different from those at the heads of the projections
of the Fermi-surface curves.
For $x=0.3$, we can observe other node lines connecting
the points of $(-\pi,0)$, $(0,-\pi)$, $(\pi,0)$, and $(0,\pi)$.
These node structures seem to be consistent with those proposed
by Usui {\it et al.} \cite{Usui}.

In Figs.~6, we show $\Delta_{\rm A}(\bm{k})$
and $\Delta_{\rm B}(\bm{k})$ for $x=0.4$ and $T=0.0355$.
The $d$-wave node structure is similar to that in Figs.~5,
but we do not observe the node lines connecting four points of
$(-\pi,0)$, $(0,-\pi)$, $(\pi,0)$, and $(0,\pi)$ in Fig.~6(b).
Rather the gap function $\Delta_{\rm B}(\bm{k})$ changes
its sign for three times on the straight line from $(\pi,0)$ to $(0,\pi)$.
The gap function $\Delta_{\rm B}(\bm{k})$ seems to includes
the components of the symmetry higher than the $d$-wave,
for instance, $g$-wave \cite{Wu}.

The gap structure on the Fermi-surface curves
around at the X and Y points has been investigated
in NdO$_{0.71}$F$_{0.29}$BiS$_2$ with the use of
an angle-resolved photoemission spectroscopy \cite{Ota}.
The large anisotropy has been observed and
the gap size on the point at which the Fermi-surface curve
crosses the line from $(\pi, 0)$ to $(\pi,\pi)$ is smaller than
that on the point at which the Fermi-surface curve
crosses the line from $(\pi, 0)$ to $(0,\pi)$.
The tendency seems to agree with the present results,
but in order to conclude the existence of the node,
it is necessary to perform further experimental and theoretical
investigations.

Next we consider the gap functions for $x >0.45$.
In Fig.~7(a) and 8(a), we depict the gap functions $\Delta_{\rm A}(\bm{k})$
for $x=0.5$ and $0.6$, respectively.
Note that for $x=0.6$, we also depict
the Fermi-surface curve originating from the A band around at the X and Y points.
In both cases, we do not find the nodes characteristic to the $d$-wave
in the gap function.
The node lines satisfying $\cos k_x +\cos k_y=0$ seem to be observed,
but it is difficult to confirm them only from this figure.

In Figs.~7(b) and 8(b), we show the gap functions $\Delta_{\rm B}(\bm{k})$
and the large Fermi-surface curves for $x=0.5$ and $0.6$, respectively.
Also in these cases, the nodes characteristic to the $d$-wave gap
cannot be observed.
Note that the structures in the gap functions are
quite similar to each other irrespective of the difference in $x$.
We find several numbers of node lines expressed as $k_y=\pm k_x+C$,
where $C$ indicates an appropriate constant.
When we change $\bm{k}$ in $\Delta_{\rm B}(\bm{k})$
along the line from the $\Gamma$ to M points,
the gap changes its sign for several times.

We remark that the Fermi-surface structure becomes different
between $x=0.5$ and $0.6$.
In particular, the curvature of the Fermi-surface curves for $x=0.5$
is larger than that for $x=0.6$.
Since the nodes are well expressed by the lines of $k_y=\pm k_x+C$,
the nodes of the gap easily appear on the Fermi-surface curves
with large curvature for $x=0.5$ in comparison with those for $x=0.6$.
From Figs.~7 and 8, there seems to occur a peculiar possibility
that the Fermi-surface curves tend to
avoid the extended $s$-wave gap functions with large amplitude.
This issue should be investigated in detail in future.

\section{Discussion and Summary}

In this paper, we have solved the gap equation to obtain the
superconducting transition temperature $T_{\rm c}$
and the gap function within the RPA
on the basis of the two-band Hubbard model
for BiS$_2$-based layered compounds.
When the topology of the Fermi-surface curves
of the B band drastically changes around at $x=0.45$,
we have observed the change of the symmetry
of the gap function.
For $x<0.45$, we have obtained the $d$-wave gap with the nodes
on the lines of $k_x=\pm k_y$, which do not cross the Fermi-surface
curves with the pocket-like disconnected shapes around at the X and Y points.
On the other hand, for $x>0.45$, we have found the extended $s$-wave gap.
An interesting issue is that the amplitude of the gap seems to be small or zero
on the Fermi-surface curves.

The effect of the nodes on the temperature dependence of
physical quantities is quite interesting
for the comparison with experimental result.
In particular, it has been reported that the temperature dependence of
the penetration depth indicates the $s$- or extended $s$-wave gap
both for $x=0.3$ and $0.5$ in NdO$_{1-x}$F$_x$BiS$_2$ \cite{Jiao}.
In the present results, we have found the $d$-wave gap for $x<0.45$,
but the main nodes of $d$-wave gap do not cross the Fermi-surface curve.
For $x>0.45$, we have obtained the extended $s$-wave gap.
However, for both gap functions, we have found the node structure
on the Fermi-surface curves depending on the doping $x$.
It is interesting to clarify the temperature dependence
of the penetration depth for the gap function with the present node structure.
This is one of future problems.

In the present calculations, we have considered only the Coulomb interactions.
However, around at the Lifshitz transition point
at which the Fermi-surface topology is abruptly changed,
the coupling between the charge fluctuations and lattice vibrations
seems to be important.
In fact, the role of phonons in BiS$_2$-based materials has been
discussed intensively \cite{Yildirim,Wan,Li}.
We believe that it is necessary to include electron-phonon interaction
to the present two-orbital Hubbard model.
It is another future problem.

In summary, we have discussed the gap symmetry of
BiS$_2$-based layered superconductors by analyzing
the two-orbital Hubbard mode within the RPA.
For $x<0.45$, we have found the $d$-wave gap on the pocket-like
disconnected Fermi-surface curves, whereas for $x>0.45$,
the extended $s$-wave gap has been found on
a couple of large Fermi-surface curves with the centers at the
$\Gamma$ and M points.
The symmetry change in the gap function is believed to
be induced by the topological change in the Fermi-surface structure.

\section*{Acknowledgement}

The authors thank Y. Aoki, K. Hattori, R. Higashinaka, K. Kubo, T. D. Matsuda,
Y. Mizuguchi, Y. Ota, and K. Ueda for useful discussions
on BiS$_2$-based layered superconductors.
The computation in this work has been partly done using the facilities of the
Supercomputer Center of Institute for Solid State Physics, University of Tokyo.





\end{document}